\documentclass[aps,twocolumn,showpacs,showkeys]{revtex4}

\begin{document}


\newcommand\beq{\begin{equation}}
\newcommand\eeq{\end{equation}}
\newcommand\beqa{\begin{eqnarray}}
\newcommand\eeqa{\end{eqnarray}}
\newcommand\ket[1]{|#1\rangle}
\newcommand\bra[1]{\langle#1|}
\newcommand\scalar[2]{\langle#1|#2\rangle}

\newcommand\jo[3]{\textbf{#1}, #3 (#2)}


\title{\Large\textbf{Secure Quantum Bit Commitment
                     Using Unstable Particles}}

\author{Chi-Yee Cheung}

\email{cheung@phys.sinica.edu.tw}

\affiliation{Institute of Physics, Academia Sinica\\
             Taipei, Taiwan 11529, Republic of China\\}


\begin{abstract}

Using unstable particles which decay by emitting neutrinos,
we propose a quantum bit commitment protocol that is
humanly impossible to break. Neutrinos carry away quantum
information, but their interaction with matter is so weak
that it would take an astronomically-sized machine just to
catch them, not to mention performing controlled unitary
operations on them. As a result quantum information is
lost, and cheating is not possible even if the participants
had access to the most powerful quantum computers that
could ever be built. Therefore, for all practical purposes,
our new protocol is as good as unconditionally secure.

\end{abstract}



\maketitle


Bit commitment is a simple cryptographic protocol involving
two parties, customarily named Alice and Bob. Alice commits
to Bob a secret bit $b\in\{0,1\}$ that is to be revealed at
some later time. In order to ensure Bob that she will keep
her commitment, Alice provides Bob with a piece of evidence
with which he can verify her honesty when she unveils. The
security of bit commitment is an important issue because it
can be used to implement other more complicated
cryptographic protocols \cite{Brassard-96}.

A bit commitment protocol is secure if it satisfies the
following two conditions. (1) Concealing: Bob cannot
determine the value of $b$ before Alice unveils it; (2)
Binding: Alice cannot change $b$ without Bob's knowledge.
Furthermore, if the protocol remains secure even if Alice
and Bob had capabilities limited only by the laws of nature
(this is sometimes referred to as the parties having
unlimited computational power), then it is said to be
unconditionally secure.

Consider a simple example. Alice writes down her bit $b$ on
a piece of paper and locks it in a box, which she gives to
Bob as evidence of her commitment. She unveils by
announcing the value of $b$ and giving the key to Bob for
verification. This protocol seems secure because Alice
cannot change the bit without access to the box, and Bob
cannot open the box without the key. However as with other
classical cryptographic schemes, it is not unconditionally
secure, because, \textit{e.g.}, Bob's ability to open the
box by himself is not in violation of any natural laws. By
introducing quantum mechanics into the bit commitment game,
one hopes to achieve unconditional security which is
guaranteed by the laws of nature. In a quantum bit
commitment (QBC) protocol, Alice and Bob execute a series
of quantum and classical operations, which results in a
quantum state with density matrix $\rho_B^{(b)}$ in Bob's
hand. If
 \beq
 \rho_B^{(0)}=\rho_B^{(1)}\label{perfect},
 \eeq
then the protocol is perfect concealing, and Bob is not
able to extract any information about the value of $b$ from
$\rho_B^{(b)}$. That means $b$ is encoded in the
representation of $\rho_B^{(b)}$. In the unveiling phase,
Alice is required to specify the representation so that Bob
can check if she is honest.

It is generally accepted that unconditionally secure
quantum bit commitment is ruled out as a matter of
principle. This is due to a 1997 no-go theorem
\cite{Lo-97,Mayers97} which states that, if Alice and Bob
have access to quantum computers, then no QBC protocol can
be concealing and binding at the same time. Furthermore, it
has been shown recently that this is the case even if Bob
employs secret parameters unknown to Alice
\cite{Cheung05,Cheung06}.

Given the fact that unconditionally secure QBC is ruled out
in theory, it does not follow, however, that all protocols
are breakable within human capabilities. This is relevant
because QBC is a cryptographic task meant to be implemented
in the real world; so if the security of a protocol is
humanly impossible to break, then practically it is as good
as unconditionally secure, even though it is not in the
mathematical sense. The purpose of this paper is to show
that the laws of physics permit a level of security which
is not jeopardized by even the most powerful possible
quantum computers.

Before proceeding further, we briefly review the original
arguments leading to the no-go result for the perfect
concealing case \cite{Lo-97,Mayers97}. (For the
near-perfect case where $\rho_B^{(0)}\approx\rho_B^{(1)}$,
see Refs.\ \cite{Lo-97,Mayers97,Cheung06}.) The crucial
ingredient is the observation that the whole commitment
process, which may involve any number of rounds of quantum
and classical exchanges between Alice and Bob, can be
represented by an unitary transformation ${\cal
U}_{AB}^{(b)}$ on some initial pure state
$\ket{\phi^{(b)}_{AB}}$. Therefore at the end of the
commitment phase, there exists a pure state
 \beq
 \ket{\Psi^{(b)}_{AB}}=\mathcal{U}_{AB}^{(b)}
 \ket{\phi^{(b)}_{AB}}
 \eeq
in the combined Hilbert space $H_A\otimes H_B$ of Alice and
Bob, instead of just a mixed state $\rho_B^{(b)}$ in $H_B$.
$\ket{\Psi^{(b)}_{AB}}$ is called a quantum purification of
$\rho^{(b)}_B$, such that
 \beq
 {\rm Tr}_A~ \ket{\Psi^{(b)}_{AB}}
 \bra{\Psi^{(b)}_{AB}}
 =\rho_B^{(b)}, \label{reduced}
 \eeq
where the trace is over Alice's share of the state. In this
approach, all undisclosed parameters are left undetermined
at the quantum level. Note that the implementation of
${\cal U}_{AB}^{(b)}$ in general requires Alice and Bob to
have access to quantum computers, which is consistent with
the assumption that they have unlimited computational
power.

The concealing condition, Eq.\ (\ref{perfect}), together
with Schmidt decomposition theorem
\cite{Hughston-93,Schmidt06}, implies that
$\ket{\Psi^{(0)}_{AB}}$ and $\ket{\Psi^{(1)}_{AB}}$ can be
written as
 \beqa
 &&\ket{\Psi^{(0)}_{AB}}=\sum_i \sqrt{\lambda^i}\,
 \ket{e^i_A}\otimes\ket{\psi^i_B},\label{Psi0}\\
 &&\ket{\Psi^{(1)}_{AB}}=\sum_i \sqrt{\lambda^i}\,
 \ket{f^i_A}\otimes\ket{\psi^i_B},\label{Psi1}
 \eeqa
where $\{\ket{e^i_A}\}$, $\{\ket{f^i_A}\}$, and
$\{\ket{\psi^i_B\}}$ are orthonormal bases in $H_A$ and
$H_B$ as indicated. Notice that $\ket{\Psi^{(0)}_{AB}}$ and
$\ket{\Psi^{(1)}_{AB}}$ are identical except for the bases
$\{\ket{e^i_A}\}$ and $\{\ket{f^i_A}\}$, which are related
by an unitary operator $U_A$:
 \beq
 \ket{f^i_A}=U_A \ket{e^i_A}.
 \label{UA-1}
 \eeq
Hence we also have
 \beq
 \ket{\Psi^{(1)}_{AB}} = U_A \ket{\Psi^{(0)}_{AB}}.
 \label{UA-2}
 \eeq
It is important to note that $U_A$ acts on $H_A$ only so
that Alice can implement it without Bob's help. It then
follows that she can cheat with the following sure-win
strategy (called EPR attack). Alice always commits to $b=0$
in the beginning. Later on if she wants to keep her initial
commitment, she simply follows the protocol honestly to the
end. Otherwise if she wants to switch to $b=1$ instead, she
only needs to apply $U_A$ to the qubits in her control, and
then proceeds as if she had committed to $b=1$ in the first
place. Bob would conclude that Alice is honest in either
case, because his density matrix $\rho_B^{(b)}$ is not
affected by the transformation $U_A$. Therefore, if a QBC
protocol is concealing, it cannot be binding at the same
time.

Notice that, in the impossibility proof outlined above, it
is implicitly assumed that Alice can maintain full control
over her share of the pure state $\ket{\Psi^{(b)}_{AB}}$
indefinitely after the end of the commitment phase. This is
however not possible if the protocol involves unstable
particles which can carry quantum information only for a
finite period of time. Consider, for example, the neutron
($n$) which decays spontaneously via weak interaction
($\beta$-decay) into a proton ($p$), an electron ($e$), and
an anti-electron neutrino ($\bar\nu_e$),
 \beq
 n\rightarrow p+e+\bar\nu_e,
 \eeq
with a mean lifetime of $\tau_n=885.7$ seconds
\cite{PDG-04}. If Alice is required to take certain action
on a neutron, it is very unlikely that she could maintain
full control over the resulting state for a period much
longer than a few $\tau_n$'s.

One might argue that, by coherent manipulation of the decay
products, it is still possible to control the spin of the
neutron after it decays. This is in principle true. However
to do so, one must be able to preserve the coherence
between the decay products and the rest of the system for
an indefinite length of time, which is practically
impossible. The reason is that the wave functions of the
light particles ($e$ and $\bar\nu_e$) propagate outward in
all directions at near light-speed $c$, so that the volume
containing the decay fragments increases with time as
$(ct)^3$, which would soon encloses the entire earth.
Moreover there will be numerous neutrons decaying into the
same volume, and one would have to be able to identify and
manipulate the wave functions originating from a single
neutron, without disturbing the others. On top of this, an
even more serious problem is that the (anti-)neutrino
participates in weak interactions only. Its interaction
with matter is so weak that a ``neutrino passing through
the entire earth has less than one chance in a thousand
billion of being stopped by terrestrial matter"
\cite{Bahcall00}. That means, on the one hand, the earth is
not likely to cause decoherence to the anti-neutrino. On
the other hand, one would need a detector a thousand
billion times the size of the earth just to catch a
particular neutrino, not to mention a machine to perform
controlled unitary transformations on it. And there are
additional complications, \textit{e.g.}, neutrinos change
identities due to flavor oscillations \cite{SNO-02}.
Certainly, by measuring the momenta of the electron and the
proton, one could determine the momentum and spin direction
of $\bar\nu_e$, without actually detecting $\bar\nu_e$
itself. However this operation is neither controlled nor
unitary, and hence is not useful to the cheating party.
From the above discussion, we conclude that, for all
practical purposes, the quantum information carried by a
neutron is lost after it decays.

Besides the neutron, there are many other naturally
occurring or artificial weakly decaying particles with
different lifetimes. For example, the muon ($\mu$) and the
Cobalt-60 nucleus ($\,^{60}\textrm{Co}$) are also unstable
against $\beta$-decay with mean lifetimes of $2.2\times
e^{-6}$ second \cite{PDG-04} and 5.3 years
\cite{Livingood-41} respectively.

In the QBC protocol to be proposed below, we shall
generically call the weakly decaying particle $W$, which
could be an elementary particle or atomic nucleus. The $W$
carries spin $J\ne 0$, and it beta decays into a daughter
particle $w$,
 \beq
 W\rightarrow w+e+\bar\nu_e,
 \eeq
with a mean lifetime $\tau_{\textsl{w}}$. For simplicity,
and without loss of generality, we shall take $J=1/2$. As
we shall see, the security of this protocol is guaranteed
by the laws of physics, independent of whether quantum
computers are available or not. Let $N$ be the security
parameter, and
 \beqa
 \ket{+\hat z}&=&\ket{0},\,\,\,\ket{-\hat z}=\ket{1},\\
 \ket{\pm\hat x}&=&\frac{1}{\sqrt{2}}
 \Big(\ket{0}\pm\ket{1}\Big).
 \eeqa
The new protocol is specified as follows.
\begin{itemize}
\item[]{\hspace{-1.3em}\textbf{Commitment phase:}}
\item[1.]{Bob sends Alice an ordered sequence of $N$ stable
qubits, each drawn independently from the set
 \beq
 \mathcal{B}=\{\ket{+\hat z},\ket{-\hat z},
 \ket{+\hat x},\ket{-\hat x}\}
 \eeq
with even probability.}
\item[2.]{To commit to $b=0$, Alice keeps the stable
qubits intact. For $b=1$, she swaps the states of the
stable qubits into $N$ unstable $W$-states, and measures
the momentum of the electron emitted from each $W$ when it
decays.}
\item[]{\hspace{-1.3em}\textbf{Unveiling phase:}}
\item[1.]{Alice unveils the value of $b$.
For $b=0$, she sends the $N$ stable qubits back to Bob in
the original order. For $b=1$, she announces the electron
data obtained previously from her measurements. To ensure
the security of the protocol, unveiling should take place
after a finite fraction of the $W$'s has theoretically
decayed.}
\item[2.]{Bob verifies Alice's honesty. If $b=0$,
he checks if the states of the stable qubits are the same
as before. If $b=1$, he calculates the electron asymmetry
using Alice's data as follows. Let $\hat e_i$ be the
polarization vector of the $i$-th $W$, where
 \beq
 \hat e_i\in\{+\hat z,-\hat z,+\hat x,-\hat x\}
 \eeq
corresponding respectively to the four states in the set
$\mathcal{B}$. $\vec p_i$ is the momentum of the electron
emitted by the $i$-th $W$, and
 \beq
 \theta_i=\textrm{cos}^{-1}
 \Big(\hat e_i\cdot\vec p_i/|\vec p_i|\Big).
 \eeq
Let $n(\theta,p)$ be the number of events for which
$\theta_i=\theta$ and $|\vec p_i|=p$, then the asymmetry
$A(\theta,p)$ is given by
 \beq
 A(\theta,p)=\frac{n(\theta,p)-n(\pi-\theta,p)}
 {n(\theta,p)+n(\pi-\theta,p)}. \label{asymmetry}
 \eeq
$A(\theta,p)$ should reproduce the known experimental
results for all $\theta$ and $p$, otherwise Alice is
cheating.}
\end{itemize}

Before proceeding to analyze the security of the protocol,
let us first explain the physics behind Eq.\
(\ref{asymmetry}). Consider a collection of $W$'s polarized
along $\hat e$. Let $\vec\sigma$ be the spin operator of
the $W$, and $\vec p$ the electron momentum, with $\hat
e\cdot\vec p/|\vec p\,|=\textrm{cos}(\theta)$. Then
$n(\theta,p)$ is a measure of the expectation value of the
operator $\vec\sigma\cdot\vec p\,$ in the decay process,
namely,
 \beq
 n(\theta,p)\propto\langle\vec\sigma\cdot\vec p\,\rangle.
 \eeq
Under parity inversion, $\vec p$ changes sign, but the spin
operator $\vec\sigma$ does not; hence $\vec\sigma\cdot\vec
p$ is a parity-odd (pseudoscalar) operator. If parity is
conserved in $\beta$-decay, we must have
 \beq
 \langle\vec\sigma\cdot\vec p\,\rangle
 =\langle\vec\sigma\cdot(-\vec p\,)\rangle,
 \eeq
which implies
 \beq
 n(\theta,p)=n(\pi-\theta,p),
 \eeq
and consequently $A(\theta,p)=0$ for all $\theta$ and $p$.
In reality, parity is maximally violated in weak
interactions which is the underlying mechanism behind
$\beta$-decay \cite{Lee-56,Wu-57}. Hence if the data
provided by Alice are genuine, Bob would find
$A(\theta,p)\ne 0$. Otherwise, if she assigned the electron
momentum $\vec p_i$ randomly by hand, then Bob would obtain
the parity conserving result of $A(\theta,p)=0$, which is a
signal of cheating.

It is trivial to prove that the protocol is concealing. Let
$\ket{\Phi_{\alpha\beta}}$ be the total state produced by
Bob, where $\alpha$ denotes the stable qubits to be sent to
Alice, and $\beta$ the ancillas if any. After sending the
$\alpha$-sector to Alice, Bob's density matrix is given by
 \beq
 \rho^{}_{\hspace{-0.1ex}
 \raisebox{-0.3ex}{$\scriptstyle\beta$}}
 =\textrm{Tr}_{\alpha}
 \ket{\Phi_{\alpha\beta}}\bra{\Phi_{\alpha\beta}}.
 \eeq
Clearly whatever Alice does to commit, the reduced density
matrix on Bob's side is unaffected. Hence at the end of the
commitment phase,
 \beq
 \rho^{(0)}_B=\rho^{(1)}_B=
 \rho^{}_{\hspace{-0.1ex}
 \raisebox{-0.3ex}{$\scriptstyle\beta$}},
 \label{perfect-2}
 \eeq
and the protocol is perfectly concealing.

Next we prove that it is binding. As explained before, the
quantum information carried by a $W$ is practically lost
after its decay. It follows that if Alice first commits to
$b=1$ and changes her mind after a finite fraction of the
$W$'s has decayed, her chance of escaping Bob's detection
is exponentially small.

The question remains, if Alice first commits to $b=0$,
could she change to $b=1$ without Bob's knowledge?
Obviously the only way to proceed is to swap the states of
the stable particles into unstable $W$'s, and wait for them
to decay. However she cannot postpone her decision until
the very last moment, because the $W$'s take time to decay.
Suppose Bob wants to bind Alice to her commitment for a
period no shorter than $T$, then the following arrangement
is sufficient, though not unique. Bob instructs Alice to
use a kind of unstable particles with mean lifetime
$\tau_{\textsl{w}}=10T$, and Alice unveils $2T$ after the
conclusion of the commitment procedure. In this situation,
if Alice commits to $b=1$ at the beginning, then by the
time she unveils the average number of $W$'s decayed is
given by
 \beq
 \delta\hspace{-0.2ex}N(2T)= N(1-e^{-0.2}).
 \eeq
However if she first commits to $b=0$, and changes her mind
at a time $T$ before unveiling, then the number of recorded
decay events would be smaller:
 \beq
 \delta\hspace{-0.2ex}N(T)=N(1-e^{-0.1}).
 \eeq
That means, to unveil $b=1$, Alice would have to
artificially generate
$\delta\hspace{-0.2ex}N(2T)-\delta\hspace{-0.2ex}N(T)\approx
N/10$ electron momentum data. It is important to note that
these artificial data contribute to the denominator but not
the numerator of Eq.\ (\ref{asymmetry}). As a result, Bob
would obtain an asymmetry which is smaller than what it
should be by a factor of
 \beq
 F=\delta\hspace{-0.2ex} N(T)/
 \delta\hspace{-0.2ex} N(2T) \approx 1/2.
 \eeq
In an ideal world where there are no systematic errors, and
statistical errors can be made as small as desired, a $1/2$
reduction in $A(\theta,p)$ is a clean and clear signal of
cheating by Alice. One can readily show that, for any
typical $(\theta,p)$, the chance of obtaining the correct
$A(\theta,p)$ by statistical fluctuation is exponentially
small for large $N$. This concludes the proof that our new
protocol is secure. We emphasize that it would remain
secure even if Alice had access to the most powerful
quantum computer that could ever be built.

In summary, we have constructed a QBC protocol where some
of the particles involved are unstable. Unstable particles
can carry quantum information only for a finite period of
time, and this property turns out to be useful in
constructing secure QBC protocols. The idea is that the
spontaneous decaying of the unstable particles may render
the associated quantum information uncontrollable. If so,
then cheating by EPR attack becomes impossible. In the case
of any weakly decaying particles emitting neutrinos,
controlling the decay products in a coherent manner would
require an astronomically-sized quantum computer operating
on neutrinos, which is clearly beyond the human capability
to build. Therefore, for all practical purposes, our
protocol is as good as unconditionally secure.


\acknowledgments {The author is grateful to C. H. Bennett
and T. Beals for useful comments.}



\end{document}